# Competition between decay and dissociation of core-excited OCS studied by X-ray scattering

M. Magnuson, J. Guo, C. Såthe, J.-E. Rubensson, and J. Nordgren
*Department of Physics, Uppsala University, Box 530, S-75121 Uppsala, Sweden*

P. Glans
*Atomic Physics, Physics Department, Stockholm University, Frescativ. 24, S-10405 Stockholm, Sweden*

L. Yang, P. Salek, and H. Ågren
*Computational Physics, Institute of Physics and Measurement Technology, Linköping University, S-58183, Linköping, Sweden*

**Abstract**

We show the first evidence of dissociation during resonant inelastic soft X-ray scattering. Carbon and oxygen *K*-shell and sulfur *L*-shell resonant and non-resonant X-ray emission spectra were measured using monochromatic synchrotron radiation for excitation and ionization. After sulfur, $L_{2,3} \rightarrow \pi^*, \sigma^*$ excitation, atomic lines are observed in the emission spectra as a consequence of competition between de-excitation and dissociation. In contrast the carbon and oxygen spectra show weaker line shape variations and no atomic lines. The spectra are compared to results from *ab initio* calculations and the discussion of the dissociation paths is based on calculated potential energy surfaces and atomic transition energies.

# 1 Introduction

The rapid advancement in synchrotron radiation instrumentation in the last few years has led to much progress in the field of molecular core-level spectroscopies. The energy selectivity, which is obtained using monochromatic synchrotron radiation, allows resonant excitation of different types of molecular states. In particular, core excitation of small molecules with repulsive intermediate states has attracted attention, for example in HBr [1], HCl [2,3,4], $O_2$ [5], $H_2S$ [6], and in small polyatomic molecules like $PH_3$ and $BF_3$ [7]. An improvement in the theoretical understanding of these processes has also been achieved lately [8,9,10,11].

In decay spectra, the dissociation is manifested by the simultaneous presence of molecular and atomic contributions. The core-excited molecule may fragment before de-excitation (dissociation channel prior), leading to atomic (or molecular radical) decay lines, or the de-excitation is followed by dissociation (de-excitation channel prior), which leads to broad molecular features. The time for dissociation and the lifetime of the core hole have pronounced effects on line positions, line widths, transition rates, and vibrational fine structures. A fast dissociation results in sharp atomic lines governed by the different dissociation channels involved, while slower dissociation results in broad molecular features. If both molecular and atomic features are observed with similar intensities, the dissociation and the core-hole decay processes occur on the same time scale, i.e. typically in the $10^{-14}$ s range.

The dissociative core-excitation processes have until now been studied in the non-radiative deexcitation channel, while no results for the radiative channel has so far been presented. The radiative channel provides new insights owing to the dipole selectivity of the process, and the implementation of local selection rules [12,13]. This provides a simple way to characterize the dissociated fragments. The lack of experimental data is due to the low fluorescence yields and instrument efficiencies associated with soft X-ray emission (SXE) in the sub-keV energy region which makes the measurements more demanding and intense synchrotron radiation (SR) sources are therefore a prerequisite.

Resonant SXE is based on a scattering process involving two photons. The process can be described in the following way;

$$A + \hbar\omega \rightarrow A(k^{-1}v) \rightarrow A(n^{-1}v) + \hbar\omega',$$

where $k$, $n$ and $v$ are the levels defined by the core, occupied valence and unoccupied valence molecular orbitals (MO's), respectively. $\omega$ and $\omega'$ are the frequencies of the incoming and outgoing photons, and $A(k^{-1}v)$ denotes the core-excited state, in which an electron from the core level $k$ has been excited to the unoccupied MO $v$. $A(n^{-1}v)$ is the valence-excited state, where the core hole $k^{-1}$ has been filled by an electron from the valence orbital $n$ through the emission process. It represents a resonant X-ray emission process if $v$ is a discrete MO, but a nonresonant process if $v$ is an orbital in the continuum. For resonant excitation, the fluorescence decay may then occur either via a spectator transition (when $n \neq v$), where an electron from one of the occupied orbitals fills the core hole while the excited electron remains as a spectator in the previously unoccupied orbital, or via a participator transition (when $n=v$), where the excited electron itself fills the core hole.

In this work carbonyl sulfide (OCS) has been studied. The OCS molecule has 30 electrons and a linear geometry in the ground state. The corresponding electronic configuration is $1\sigma^2 2\sigma^2 3\sigma^2 4\sigma^2 5\sigma^2 1\pi^4 6\sigma^2 7\sigma^2 8\sigma^2 2\pi^4 9\sigma^2 3\pi^4$. The $1\sigma$, $2\sigma$, $3\sigma$, and $4\sigma$ orbitals correspond to the S 1s, O 1s, C 1s, and S 2s core levels, respectively, whereas the spin-orbit and molecular-field split S 2p levels are composed of the $5\sigma$ and $1\pi$ molecular orbitals. The remaining orbitals are associated with valence levels of mixed atomic character. The OCS molecule belongs to the $C_{\infty,v}$ point group and it has three vibrational modes, $v_1$, $v_2$ and $v_3$. These modes can be characterized as being due to C-S stretching, O-C-S bending, and C-O stretching, respectively. Bending of the molecule lowers the symmetry to the $C_s$ group.

We present high-resolution resonant and nonresonant SXE spectra of carbonyl sulfide near the sulfur $L_{2,3}$ absorption thresholds to demonstrate a first case of fragmentation in the X-ray scattering process [14]. Recently, it has been discussed whether the lifetime of the sulfur $L_{2,3} \rightarrow \pi^*$ core-excited states in OCS is long enough to allow a dissociation process to occur before the decay [15,16]. Calculated potential energy surfaces and atomic transition energies are used for the identification of the observed atomic lines and for an understanding of the dissociation dynamics. In addition to the sulfur spectra, we present carbon and oxygen $K$-shell SXE data. The experimental spectra are compared with results from *ab initio* calculations.

## 2  Experimental Details

The experiments were performed at beamline 7.0 at the *Advanced Light Source*, Lawrence Berkeley National Laboratory. The beamline comprises a 99-pole, 5-cm period undulator and a spherical-grating monochromator [17] covering the spectral energy range between 60-1300 eV.

The SXE spectra were recorded using a high-resolution grazing-incidence SXE spectrometer [18,19]. The spectrometer was mounted parallel to the polarization vector of the incident photon beam with its entrance slit oriented parallel to the direction of the incident beam. The pressure in the gas cell was optimized to about 2 mbar for a maximum absorption of photons in the interaction region. The interaction region was viewed by the X-ray spectrometer through a 160 nm thick polyimide window, supported by a polyimide grid and coated with 30 nm of aluminum nitride. During the SXE measurements, the incident photon beam entered the gas cell through a small pinhole of 100 μm diameter and the pressure in the experimental chamber was about $1 \times 10^{-7}$ Torr due to the outgassing from the pinhole. With 0.30 eV resolution of the monochromator of the beamline, a near-edge X-ray absorption (NEXAFS) spectrum of sulfur was measured using the photocurrent from an electrode situated inside the gas cell where the incident photon beam entered the gas cell through a 100 nm thick silicon nitride window. The spectra were normalized to the incident photon flux using a gold mesh in front of the gas cell. The bandwidth of the synchrotron light during the sulfur, carbon, and oxygen emission measurements was set to 0.2 eV, 0.7 eV, and 1.1 eV, respectively. The resolution of the X-ray emission spectrometer is estimated to be 0.2 eV, 0.7 eV and 0.8 eV, respectively.

# 3 Calculational Details

The potential energy surfaces with respect to the S $2p$ excitation/ionization, in the linear and bent geometries, were obtained with the program package DALTON [20] at the multiconfiguration self-consistent-field (MCSCF) level of theory. Calculations were performed in the $C_{2v}$ symmetry for linear geometries and in $C_s$ when the investigation included a bent conformation. In the former case ($C_{2v}$ symmetry) the inactive space was 7/1/1/0 and the active space – 4/4/4/0. In the latter one, we have 8/1 inactive orbitals, 6/3 orbitals in RAS2 space and 2/1 orbitals in RAS3. The number of electrons in RAS3 space is restricted to be between 0 and 2. A standard double zeta cc-pVDZ basis set was used.

Geometry optimizations for the $L_{2,3}$ core-ionized states were carried out, and these states were found to have linear conformations, but with the CO and CS bond lengths being slightly different from those of the ground state. The bond length between C and O is about 0.07 a.u. shorter in the ionized state (2.138 a.u.) compared to the ground state (2.205 a.u.), while the C-S bond length is about the same amount longer in the ionized state (3.075 a.u.) compared to the ground state (3.005 a.u.). The core-excited states were investigated first for the linear geometry ($C_{2v}$ symmetry group). Potential surfaces for $\pi^*$ core-excited states were also computed for various bent geometries.

Theoretical simulations of resonant and nonresonant X-ray emission as well as X-ray absorption were carried out employing the so-called static-exchange (STEX) technique. The "cc-PVTZ" type of basis sets were used for the O, C and S calculations [21]. An augmented basis set containing a large number of diffuse functions was further added in the S $L$-edge absorption calculations. For general descriptions of the theory and calculations we refer to references[22, 23, 24, 25, 26].

The theoretical analysis of the sulfur $L_{2,3}$ absorption spectrum has some special features owing to the large spin-orbit splitting and (small) molecular field splitting of the S $2p$ level. Roughly, this means a duplication of the spectra, 1.2 eV apart with a 2-1 "statistical ratio" for the $2p_{3/2}$ and $2p_{1/2}$ derived intensities. For fine structure, especially at higher energies, an intermediate coupling analysis is required, as recently proven by Fink *et al.* [27] for the Cl $2p$ NEXAFS spectrum of HCl. The current implementation of the STEX technique is based on non-degenerate core levels, and we made this assumption for the $2p$ hole of OCS, because of the small molecular-field splitting. The molecular field split, σ and π, STEX spectra were distributed into "$2p_{1/2}$" and "$2p_{3/2}$" spectra assuming the electro-static coupling limit.

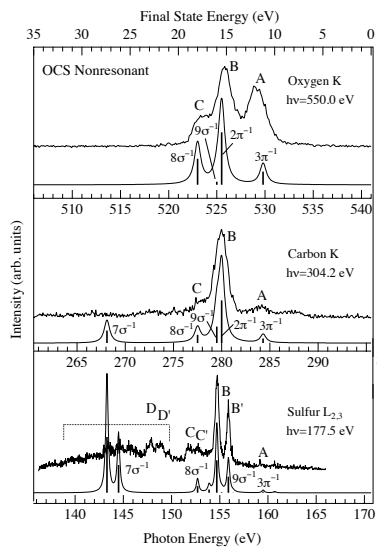

**Figure 1:** Experimental and calculated nonresonant X-ray emission spectra of oxygen, carbon and sulfur in OCS.

Atomic energy levels for the neutral sulfur $2p^5 3s^2 3p^5$ core-excited states relative to the $2p^6 3s^2 3p^4$ ground state were calculated using the HFR (Hartree-Fock with Relativistic corrections) code by Cowan[28] with the radial electrostatic integrals scaled to 80%. Energies for the $2p^5 3s^2 3p^5 \rightarrow 2p^6 3s^1 3p^5$ transitions were obtained by using optical data for the final states [29], which are significantly affected by configuration interaction. The transition energies were used in the line identification and are included in Table II. The quality of the derived energies was checked by obtaining transition energies in the same way for the isoelectronic $Ar^{2+}$ system and comparing those values to known experimental energies [30]. The agreement was found to be satisfactory with errors of less than 0.4 eV.

## 4 Results and Discussion

### 4.1 Non-resonant X-ray emission spectra

Previous, related investigations of the electronic structure of the carbonyl sulfide (OCS) molecule include photoelectron spectroscopy of the inner and outer valence region [31,32,33] and Auger spectroscopy [34,35]. In these studies extensive structures were observed in the inner valence energy region which were attributed to multielectron transitions. Electron-energy-loss spectra at the oxygen $K$, carbon $K$ and the sulfur $L_{2,3}$ edges have also been measured [36,37]. Perera and LaVilla [38] and Miyano *et al.* [39] studied the sulfur $K$ SXE of OCS. The only SXE measurements (nonresonant) at the sulfur $L$-edge so far were made by Mazalov *et al.* [40]. However, despite the lack of selectively excited SXE spectra, theoretical predictions have been made [41].

Figure 1 shows nonresonant SXE spectra of molecular OCS measured with excitation energies slightly above the oxygen and carbon $K$-edges and the sulfur $L$-edge. The final state energy scale at the top of the figure has been obtained by subtracting the SXE energy scale from the core-hole binding energies; 540.3 eV, 295.2 eV and 170.6 eV, for the oxygen $K$, carbon $K$ and sulfur $L_3$-edge, respectively. Calculated intensities using the one-center approximation are also shown for comparison. The theoretical intensities are presented as vertical bars and Lorentzian peaks positioned at the vertical binding energies of the final states, which are taken from photoelectron spectra [31,32,33]. The oxygen $K$ spectrum shows a three-peak structure denoted A, B, and C.

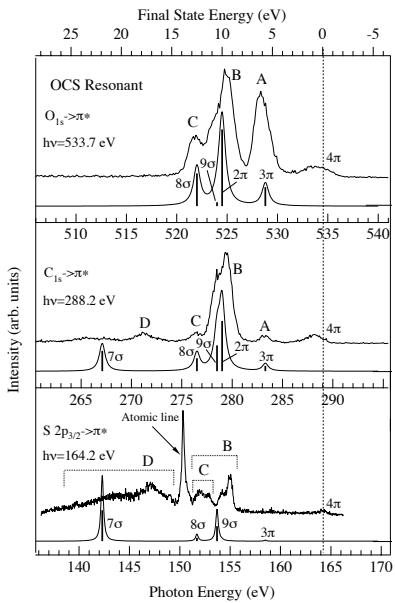

**Figure 2:** Experimental and calculated resonant X-ray emission spectra of oxygen, carbon, and sulfur. The emission peak at zero energy in each spectrum corresponds to participator transitions.

Peak A corresponds to 3π→O 1s transitions, peak B to 2π→ O 1s and 9σ→ O 1s transitions, while peak C corresponds to 8σ→ O 1s transitions. According to the calculation, the $2\pi^{-1}$ final states dominates over the 9σ at peak B.

In the carbon *K* spectrum, only feature B remains intense, while the intensity of bands A and C become much weaker than in the oxygen *K* spectrum. In addition, a weak feature appears at about 287.5 eV photon energy probably due to initial-state shake-up satellite transitions or double excitations which typically end up in this energy region. From the calculation of the carbon *K* spectrum, it is also predicted that the transition from the 9σ MO which is very weak in the oxygen *K* spectrum becomes competitive in intensity with that from the 2π MO. The difference in intensity distribution reflects the atomic composition of the molecular orbitals. In the oxygen and carbon spectra basically the *p* character is observed which makes the 7σ final state configurations which has 2*s*-character practically dipole forbidden for oxygen. Thus, the chemical C-O electron-pair bond is mainly formed by the 2π MO's.

Due to the $L_{2,3}$ spin-orbit splitting, all the emission features in the S 2*p* spectrum are split up into double structures. A spin-orbit splitting of 1.2 eV has previously been observed in core-level photoemission experiments[31, 36]. Feature A corresponding to 3π→ S $2p_{3/2,1/2}$ transitions has practically no intensity while the double peak structure denoted B and B', corresponding to the 9σ→$2p_{3/2,1/2}$ transitions, is very strong. Peaks C and C' correspond to the 8σ→$2p_{3/2,1/2}$ transitions. The 7σ→$2p_{3/2,1/2}$ transitions, which are presented as rather sharp features in the calculated spectrum are much broader in the experimental spectrum. This is a clear sign of the breakdown of the molecular orbital picture in the inner valence region, where there is a high density of overlapping σ states due to correlation state splittings [42, 43] in addition to vibrational broadening.

## 4.2 Resonant X-ray Emission Spectra

Figure 2 shows resonant SXE spectra obtained by excitations from the O 1s, C 1s, and S $2p_{3/2}$ orbitals to the $\pi^*$ orbital, which is the lowest unoccupied orbital. All the resonant peaks are slightly shifted towards lower photon energies in comparison to the nonresonant case due to the difference in the screening between the core-excited and the final states. The oxygen spectrum shows a three-peak structure similar to the nonresonant spectrum. In addition, a recombination peak is now observed at near zero energy corresponding to participator transitions in which the excited electron fills the core hole. The carbon spectrum shows a similar peak structure (denoted A, B and C) as the oxygen spectrum, although bands A and C are much weaker than band B like in the nonresonant case. In addition, a weak feature (denoted D) corresponds to the 7σ→ C 1s transitions. This feature is broadened due to configuration interaction in the inner valence region, as mentioned previously. In the resonant sulfur spectrum, the shapes of peaks B, C and D are dramatically changed compared to the nonresonant case due to a substantial vibrational broadening. However, the most conspicuous feature in the sulfur spectrum is the narrow and intense peak at 150.2 eV photon energy, which we assign to a transition in neutral atomic sulfur as will be discussed in section D.

## 4.3  S 2*p* X-ray absorption spectrum

Figure 3 shows a NEXAFS spectrum measured in the region of the S $L_{2,3}$-edges used as a reference for the SXE measurements. The spectrum consists of three strong resonance peaks followed by a weaker Rydberg series, and a shape resonance structure in the continuum, just above the ionization thresholds, presumably with multielectron excitations superimposed. The positions of the indicated $L_{2,3}$-edges are based on experimental photoelectron data [31]. Calculations of the excitations from both the $L_{2,3}$-edges are also shown for comparison. The calculated $\pi^*$ and $\sigma^*$ peaks have been shifted -0.5 eV to account for screening effects which are not sufficiently taken into account in the STEX calculations. The calculated spectrum for the $L_2$ excitation was shifted from that of the $L_3$ channel according to the experimental value of the $L_{2,3}$ spin-orbit splitting of 1.2 eV obtained from photoemission [36]. Table I shows the spectroscopic assignment which was made by using population analysis of the molecular orbital character in comparison to configuration interaction calculations by Brems *et al.* [15]. The calculated spectra from the two separate channels were added to give the total spectrum. The first peak in the total NEXAFS spectrum is solely excitation from the $2p_{3/2}(L_3)$ channel which is also predicted for peak number 5, while the remaining peaks are mixtures of excitations from both channels (see table I). Peak number 6 has contributions from many Rydberg levels converging to the ionization potential, which group together in the convoluted spectrum. As mentioned earlier, the fine structure of the Rydberg energy region is complicated due to the role of intermediate coupling. The 2*p* spin-orbit splitting creates a double-peak structure in the sulfur SXE spectra of the OCS molecule when the excitation is tuned to the mixed states (such as peaks 2,3 and 4) or the shape resonance.

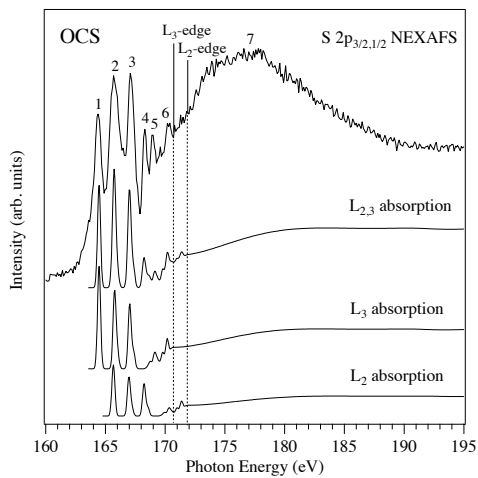

**Figure 3:** Experimental and calculated NEXAFS spectra at the sulfur $L_{2,3}$-edges used as references for the emission spectra.

## 4.4  S 2*p* X-ray emission spectra

Figure 4 shows a series of experimental sulfur $L_{2,3}$ SXE spectra on a photon energy scale. The spectra were resonantly excited at the absorption peaks, denoted by the numbers 1 through 7 in Fig. 3. The spectral shape changes substantially when going from the nonresonant and Rydberg excited topmost four spectra to the $\pi^*$ and $\sigma^*$ excited spectra. In the latter spectra the emission peaks are substantially more vibrationally broadened and the peaks also appear at somewhat lower energy due to the screening effect of the spectator electron. The larger screening resulting from a spectator electron in the first unoccupied MO's ($\pi^*$ or $\sigma^*$) than from an electron in a Rydberg orbital is most apparent in the third spectrum from the bottom (excited at 166.9 eV). As shown in Fig. 3 and Table I, this excitation energy corresponds to two transitions, $2p_{3/2} \to 4s$ and $2p_{1/2} \to \sigma^*$. In the emission spectrum the $2p_{3/2}^{-1} 4s \to 9\sigma^{-1} 4s$ transition results in a narrow peak with a small spectator shift, whereas $2p_{1/2}^{-1} \sigma^* \to 9\sigma^{-1} \sigma^*$ transition yields a broader feature with a larger spectator shift.

As was shown in Fig. 2, the resonant sulfur SXE spectrum at 164.2 eV excitation energy shows a feature without counterpart in the oxygen $K$ and carbon $K$ spectra; a narrow intense atomic line at a photon energy of 150.2 eV. This feature also shows up when the incident photons are tuned to the second absorption peak at 165.4 eV, whereas another atomic line at 0.4 eV higher emission energy (150.6 eV) is observed when exciting to the third absorption peak at 166.9 eV. The atomic lines completely disappears at the higher excitation energies where the incident photon energy is tuned to Rydberg states. The atomic lines have a width of 0.26 eV, which can be attributed to the instrumental spectrometer resolution (of about 0.2 eV) and the natural lifetime broadening (20-40 meV) of the sulfur $L_{2,3}$ core-hole [44]. As mentioned previously, the $9\sigma \to L_{2,3}$ double-peak feature at about 155 eV photon energy reflects the spin-orbit splitting of the $L$-shell excitation. As shown in Fig. 3 and Table 1, peaks No. 2, 3, 4, 6, and 7 (165.4, 166.9, 168.1, 169.9 and 177.5 eV) are due to excitations from both the $L_3$ and $L_2$ core levels. In these cases the filling up of the core levels by an electron from a single valence orbital results in two peaks. However, for the fifth absorption feature (at 168.8 eV) the calculations predict that only excitations from the $L_3$ core level contribute. This assignment is confirmed by the experimental SXE spectrum recorded at 168.8 eV, in which essentially only the $9\sigma \to L_3$ peak emerges around 155 eV. Next, we will discuss the potential energy surfaces and the dissociation.

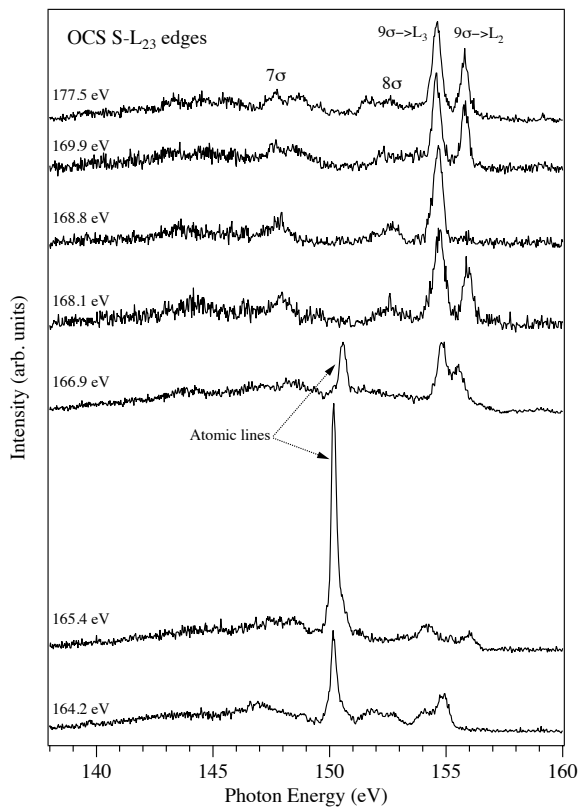

**Figure 4:** A series of X-ray emission spectra of the S $L_{2,3}$ excitations of OCS.

**Table 1:** Assignment of the S $L_{2,3}$-edge X-ray absorption spectrum of the OCS molecule [15].

| Feature | Energy (eV) | Assignment |
|---|---|---|
| 1 | 164.2 | $2p_{3/2} \to \pi^*$ |
| 2 | 165.4 | $2p_{1/2} \to \pi^*; 2p_{3/2} \to \sigma^*$ |
| 3 | 166.9 | $2p_{3/2} \to 4s; 2p_{1/2} \to \sigma^*$ |
| 4 | 168.1 | $2p_{3/2} \to 4p; 2p_{1/2} \to 4s$ |
| 5 | 168.8 | $2p_{3/2} \to 3d; 5s$ |
| 6 | 169.9 | $2p_{3/2} \to 5d; 6p; 2p_{1/2} \to 3d; 5s$ |
| 7 | 177.5 | *Shape−resonance* |

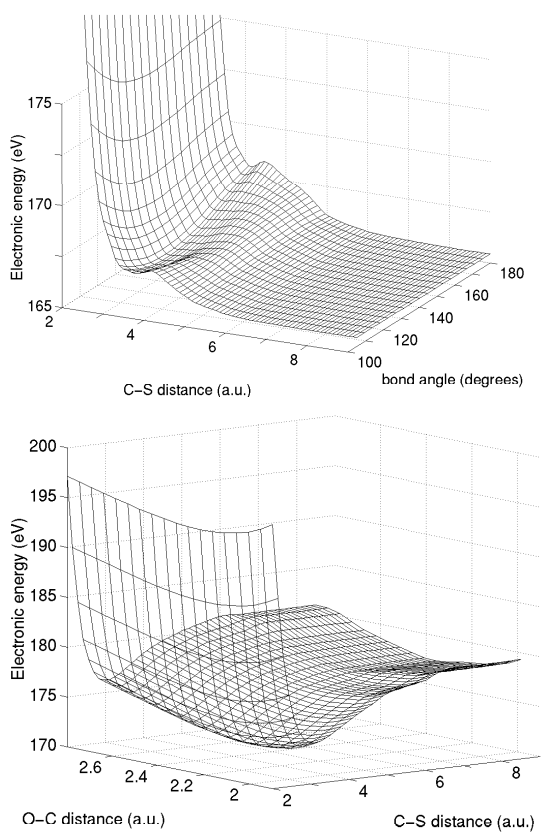

**Figure 5:** Calculated potential surfaces of the OCS molecule. a) S $2p_\sigma - \pi^*$ state two-dimensional stretch-bend surface. b) $2p_\sigma^{-1}$ state surface with respect to OC and CS bond lengths, and with fixed 180 degrees bond angle.

Figures 5a and 5b show calculated potential energy surfaces for the $2p_\sigma^{-1}\pi^*$ core-excited and the $2p_\sigma^{-1}$ core-ionized states. The core-ionized state is bound, with a potential energy minimum at a C-S distance of 3.075 a.u., and, owing to the nonbonding character of Rydberg orbitals, one can assume that this potential energy surface represents the core-to-Rydberg-excited states as well. In Fig. 5a the potential energy surface of the $2p_\sigma^{-1}\pi^*$ core-excited state is shown for various bending angles. The potential surface shows a barrier against dissociation for the linear case, i.e., at 180 degrees. This energy barrier results probably from a conical intersection and is found to increase much in calculations with a poorer account of electron correlation. The potential energy (and the barrier) decreases strongly from the 180 degrees linear geometry so that the dissociation may take place through an OCS bending vibrational mode.

It has been shown from angle-resolved ion-yield experiments that the OCS molecule indeed induces the bending vibrational mode for the $\pi^*$ excitations and that the dominating dissociation mechanism involves the breakup of the OC-S bond and the production of CO and S fragments at an angle of about 130° [15,16]. Thus, when an electron is placed in the $\pi^*$ orbital, the potential energy is lowered by the bending motion, which can be understood as a Renner-Teller vibronic coupling effect [45,46].

Figure 6 shows a set of calculated potential curves for different core-excited states in OCS, ignoring the spin-orbit splittings. It clearly shows that there is a potential energy barrier for $\pi^*$ excitations, whereas for $\sigma^*$ excitations the curves are directly dissociative. We note that the $\sigma^*$ is higher in energy at the ground state equilibrium geometry (3.005 au), and that the two curves cross at about 3.3 au.

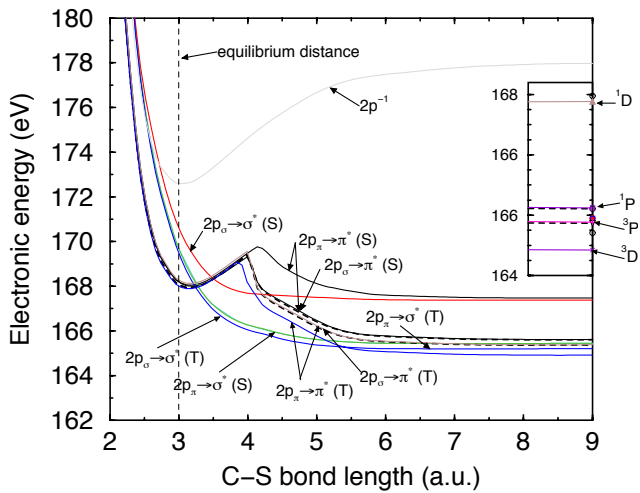

**Figure 6:** Computed potential curves for some 2p core-ionized and core excited states of OCS (ignoring the spin-orbit splitting).

The atomic lines in the resonant SXE spectra can be considered as the result of a competition between the dissociation and the decay channels. Comparing the intensity ratio of the atomic line to the molecular background in the three lowest SXE spectra shown in Fig. 4, one can conclude the presence of dissociation for all three, but to different extents. The dissociation time, $t_D$, can be estimated with the simple formula; $t_D = -\tau \ln(1-n)$, where $\tau$ is the core-hole lifetime and $n$ is the 'molecular fraction'[47]. Here, a larger molecular fraction corresponds to a longer dissociation time and we assume the SXE yield to be the same for the molecule and the fragment. The first spectrum (164.2 eV), $L_3 \to \pi^*$ excitation, shows an atomic peak that contributes to about 15% of the total intensity. If we assume the core-hole lifetime to be 22 fs (30 meV core-level width), a dissociation time of ~ 40 fs is found. In the second spectrum (165.4 eV) where there is a mixture of excitations from both the $L_2 \to \pi^*$ and $L_3 \to \sigma^*$ channels, the molecular fraction is smaller than for the first spectrum. This indicates a steeper potential path for $\sigma^*$ than for $\pi^*$ dissociation. Thus, the dissociation time is also smaller, ~ 30 fs. In the third spectrum (166.9 eV), the atomic contribution is still significant, however, in this case dissociation only occur from the $L_2^{-1}\sigma^*$ intermediate state, and the $L_3^{-1}4s$, which is also excited, only result in a molecular contribution. The estimated dissociation time is about the same as in the first spectrum (~ 40 fs). At higher excitation energies, i.e., for ionization as well as core excitation to Rydberg states, the atomic contribution completely disappears in agreement with what was expected from the calculated, bound potential energy surface in Fig. 5b.

In Table II calculated $2p^5 3s^2 3p^5 \to 2p^6 3s 3p^5$ transition energies are given. Based on this table, the strong atomic line, at 150.2 eV, in the first and second spectra is assigned to the $^3D_3 \to {}^3P_2$ transition. In the third spectrum, the atomic line is observed at a different energy, at 150.6 eV, which implies that in this case the dissociation channel is different. We assign this line to $^3D_2 \rightarrow {}^3P_{1,2}$ transitions. This line might also contribute weakly in the first and second spectra, giving rise to the high-energy shoulder of the atomic line at 150.2 eV. It should be noted that a $2p^5 3s^2 3p^5$ configuration with an $L_2$ hole cannot couple to a $^3D_3$ state, which probably explain why the atomic line at 150.2 eV is absent in the third spectrum. However, the dissociation path is complex with many near-degenerate molecular field and spin-orbit coupling split potential energy surfaces available, which also are subject to Renner-Teller effects and avoided crossings. The remarkable simplicity of the outcome in terms of one strong atomic X-ray emission line of the atomic fragment indicates strong coupling and fast interstate relaxation during the course of the dissociation.

**Table 2:** Calculated neutral atomic sulfur $L_{2,3}$ X-ray emission lines. The final state terms were taken from optical absorption data; $^3P_2$ =8.93 eV, $^3P_1$ =8.97 eV, $^3P_0$ =9.00 eV and $^1P_1$ =10.10 eV [29].

| Transition | Energy (eV) | Transition | Energy (eV) |
|---|---|---|---|
| $^1P_1 \rightarrow {}^1P_1$ | 149.01 | $^3D_1 \rightarrow {}^3P_2$ | 151.65 |
| $^3D_3 \rightarrow {}^3P_2$ | 150.27 | $^3P_2 \rightarrow {}^3P_1$ | 151.67 |
| $^3D_2 \rightarrow {}^3P_1$ | 150.80 | $^3P_2 \rightarrow {}^3P_2$ | 151.71 |
| $^3D_2 \rightarrow {}^3P_2$ | 150.84 | $^1D_2 \rightarrow {}^1P_1$ | 151.84 |
| $^3S_1 \rightarrow {}^3P_0$ | 151.07 | $^3P_0 \rightarrow {}^3P_1$ | 152.16 |
| $^3S_1 \rightarrow {}^3P_1$ | 151.10 | $^3P_1 \rightarrow {}^3P_0$ | 152.40 |
| $^3S_1 \rightarrow {}^3P_2$ | 151.14 | $^3P_1 \rightarrow {}^3P_1$ | 152.43 |
| $^3D_1 \rightarrow {}^3P_0$ | 151.58 | $^3P_1 \rightarrow {}^3P_2$ | 152.47 |
| $^3D_1 \rightarrow {}^3P_1$ | 151.61 | $^1S_0 \rightarrow {}^1P_1$ | 153.04 |

# 5 Summary

We show the first evidence of molecular dissociation prior to de-excitation in resonant soft X-ray emission. Resonant and nonresonant *K*-shell X-ray emission spectra for carbon and oxygen, and *L*-shell X-ray emission spectra for sulfur in the carbonyl sulfide molecule were measured by the use of monochromatic synchrotron radiation. In the case of S *L*-shell excitation to the $\pi^*$ and $\sigma^*$ orbitals a strong competition is observed between de-excitation and dissociation resulting in atomic-like lines in the SXE spectra. The line shapes of the molecular features in those spectra are also strongly affected by the dissociative character of the core-excited states. We find that $L_3 \rightarrow \pi^*$, $\sigma^*$ excitations lead to dissociation with the sulfur atom in a neutral $^3D_3$ core-excited state, whereas $L_2 \rightarrow \sigma^*$ excitation leads to a $^3D_2$ core-excited state. We also find evidence for a faster dissociation when exciting to $\sigma^*$ than to $\pi^*$. The slower dissociation in the latter case is supported by calculations, which give a potential energy surface for the $\pi^*$ core-excited states with a potential barrier in the linear geometry that vanishes as the molecule bends whereas the $\sigma^*$ core-excited states are dissociative also in the linear geometry.

## Acknowledgments

This work was supported by the Swedish Natural Science Research Council (NFR), the Göran Gustavsson Foundation for Research in Natural Sciences and Medicine and the Swedish Institute (SI). The experimental work at ALS, Lawrence Berkeley National Laboratory was supported by the director, Office of Energy Research, Office of Basic Energy Sciences, Materials Sciences Division of the U. S. Department of Energy, under contract No. DE-AC03-76SF00098. We thank Faris Gel'mukhanov and Reinhold Fink for rewarding discussions.